# Toward a Theory of Justice for Artificial Intelligence

## *Iason Gabriel*

*This essay explores the relationship between artificial intelligence and principles of distributive justice. Drawing upon the political philosophy of John Rawls, it holds that the basic structure of society should be understood as a composite of sociotechnical systems, and that the operation of these systems is increasingly shaped and influenced by AI. Consequently, egalitarian norms of justice apply to the technology when it is deployed in these contexts. These norms entail that the relevant AI systems must meet a certain standard of public justification, support citizens' rights, and promote substantively fair outcomes, something that requires particular attention to the impact they have on the worst-off members of society.*

Calls for justice in the context of artificial intelligence sound increasingly loud. Indeed, communications scholar Matthew Le Bui and gender studies scholar Safiya Umoja Noble have argued that we are missing a moral framework of justice altogether when it comes to evaluating the practices that constitute artificial intelligence.[1] The demand for justice represents both a need felt among those impacted by AI systems and a source of important philosophical insight. Among other things, it reframes much of the discussion around "AI ethics" by drawing attention to the fact that the moral properties of algorithms are not internal to the models themselves but rather a product of the social systems within which they are deployed. At the same time, those who want to evaluate emergent practices through the lens of justice rapidly encounter an obstacle: namely, that political theory – which is the body of thought we might hope to rely on to address these questions – has not adequately addressed technology in general, struggling to navigate a path between relative neglect and determinism. As a consequence, it is not necessarily well-equipped to speak to the role of technology in public life, let alone say something meaningful about justice and AI systems.

Taking these points in turn, much of contemporary political philosophy brackets out technological considerations, treating them as exogenous to the fundamental questions of political life. This view is found in the work of philosopher John Rawls, whose seminal work *A Theory of Justice* mentions *technology* on just three occasions. Moreover, although his account of justice appears to be for a society that has a specific sociotechnical character (that is, one with a functioning legal system, economic division of labor, capacity for taxation, and so on), knowledge about the level of technology that a society has achieved is excluded from the original position when selecting principles of justice. It is only when making a final assessment of what justice requires in specific contexts that we need to "take into account economic efficiency and the requirements of organization and technology."[2]

By contrast, technology plays a central role in Marxist thought. However, the account provided leaves little room for human choices or moral direction. Indeed, while the character of Marx's historical materialism remains subject to deep exegetical disagreement, one prominent interpretation holds that, for any given moment, the development of productive forces (that is, technology and labor) explains the nature of the mode of production (understood as the prevailing economic relations), which then shapes society's ideological superstructure, including its laws and system of beliefs.[3] Understood in this way, the development of technology still functions primarily as an exogenous force. Moreover, if prevalent moral norms are largely explained by material circumstances (and potentially nothing more than a "bourgeois ideology" in a late capitalist society), then they appear deeply, and perhaps terminally, compromised as a vantage point from which to make an independent moral evaluation.

Taken together, these accounts matter because they suggest that calls for justice in the context of AI are essentially misplaced. Understood primarily as a new technology, AI either falls outside the scope of justice or is part of a dynamic that prefigures robust moral evaluation. In this essay, I defend a different approach, one that makes claims about AI, justice, and injustice entirely appropriate.[4] This approach begins by noting that the interaction between humans and technology is a two-way process. On the one hand, we are profoundly affected by the technologies we adopt. In modern societies, technology helps to facilitate control from a single center, maintain larger organizational units, promote economic specialization, determine the meaning of authority and expertise, and shape the goals, aspirations, and self-understanding of citizens. On the other hand, we are not only acted upon by technologies, but we also create them through a process of design, experimentation, development, iteration, and adoption. Clearly, the power to shape and influence the path of technological change is not distributed evenly across society.[5] Nonetheless, choices about the content and character of new technologies are being made.

Taken together, what emerges therefore is a class of profound societal effects induced by technological change alongside a set of technological choices that shape the path of innovation via the decisions of individual technologists, markets, governance structures, and social norms. These decisions, and the institutional practices they support, compose an important subject for moral evaluation and can be assessed from the standpoint of distributive justice.

A key element of liberal political theory, as articulated by Rawls, is the distinction between the "basic structure" of society, which is subject to principles of distributive justice, and other domains of life that are not directly subject to these principles. The basic structure encompasses

> the way in which the major social institutions fit together into one system, and how they assign fundamental rights and duties and shape the division of advantages that arise through social cooperation. Thus the political constitution, the legally recognized forms of property, and the organization of the economy, and the nature of the family, all belong to the basic structure.[6]

These practices need to be structured in accordance with a common set of rules. Outside of these contexts, people are left relatively free to pursue their personal objectives, something that is important for a pluralistic society in which people have divergent goals and aspirations.

Against this backdrop, I wish to advance two claims. The first is that the basic structure of society is best understood as a composite of sociotechnical systems: that is, systems that are constituted through the interaction of human and technological elements. The claim here is not only that the basic structure contains social and technical elements, but also that these elements interact dynamically to constitute new forms of stable institutional practice and behavior.[7] The second is that AI increasingly shapes elements of the basic structure in relevant ways, and hence that its design, development, and deployment all potentially interface with principles of justice in this context.

The growing role played by AI in the operation of key institutions and practices is well illustrated by the criminal justice system, in which risk-assessment algorithms increasingly determine a person's eligibility for bail or parole, facial recognition technology has been used to augment police capabilities, and AI systems direct the allocation of policing resources using predictive analytics. In the context of economic mobility and access to key public services such as welfare provision, the use of algorithmic tools is similarly influential, determining who is eligible for welfare support, who has access to public housing, and which families are engaged by child services.[8] Meanwhile, in the economic sphere, financial institutions use these models to determine who has access to loans, mortgages, and insurance. Finally, these tools have a wider impact on the economic prospects of citizens via their integration into job recommendation search engines – helping to determine who is shown what opportunities – and via the tools used by educational institutions to allocate students or advertise opportunities for higher education.[9]

In each case, AI is not simply an additional ingredient that supervenes onto a stable practice leaving the fundamental elements of that practice untouched. Rather, AI interacts with the behavior of human decision-makers to shape the character of these practices, including how they distribute benefits and burdens across the population. In the context of criminal justice, for example, there is significant concern that parole recommendation algorithms compound historical injustice by recreating and extending racial bias found in the training data.[10] In the context of government services, AI has changed the nature of welfare provision, including who can access it and on what terms, with political scientist Virginia Eubanks documenting the emergence of a "feedback loop of injustice" whereby "marginalized groups face higher levels of data collection when they access public benefits … [which] acts to reinforce their marginality when it is used to target them for suspicion and extra scrutiny."[11] Meanwhile, in the domain of credit scoring and access to financial services, legal scholar Frank Pasquale has raised concerns about the increasingly significant role played by a person's algorithmically determined "digital reputation" as a major determinant of their life chances.[12] Speaking to the dynamic interaction between these systems and the social environment in which they are deployed, Pasquale notes that "unlike the engineer, whose studies do nothing to the bridges she examines, a credit scoring system increases the chance of a consumer defaulting once it labels him a risk and prices a loan accordingly."[13] Given the potential serious knock-on effects these practices have for equality at the societal level, they have driven concerns about "digital redlining" – with entire groups of people encountering new barriers to opportunity – and the emergence of what, with respect to race, sociologist Ruha Benjamin terms "the New Jim Code."[14]

To be clear, the concerns that arise in these contexts are not only concerns about distributive justice, they also involve racial justice, criminal justice, historic injustice, and the disciplinary power of institutions.[15] However, principles of distributive justice that spell out how major institutions ought to allocate opportunities and resources are also relevant here. Moreover, they can help explain what is morally problematic about these practices and show how these harms can be addressed.

According to the Rawlsian framework, there are two key grounds that make a practice subject to regulation by principles of distributive justice, both of which are now met by the aforementioned AI systems. First, these principles apply to institutions that are necessary in order to maintain "background justice" over time.[16] According to this view, a social practice should be regulated by principles of distributive justice when, without this intervention, the compound effect of individual choices would lead to forms of inequality that threaten the equal standing and autonomy of citizens.[17] For example, the uninterrupted interplay of market forces would likely leave some people so badly off that they could no longer give meaningful consent to the institutional practices that structure their lives, and would instead have to accept whatever arrangement was offered to them by the rich and powerful. To avoid this outcome, the practices that make up the basic structure need to be regulated in ways that support background justice, counteracting the tendency of multiple individual transactions to distort the distribution of income and wealth over time.

What is important for our purpose is that in modern societies, background justice is increasingly mediated algorithmically. Across various contexts, including social service provision, credit allocation, and insurance eligibility decisions, AI systems have now taken on this critical function. By making assessments or predictions based upon an individual's past choices, and by providing decisions or recommendations that then shape that person's opportunity set, these systems exert a strong influence on the unfolding relationship between individual choices and collective outcomes. Moreover, unless their operation is aligned with principles of distributive justice, these systems could compound inequality in ways that a just society aims to forestall.

Second, principles of distributive justice apply to certain practices because they exercise a "profound and pervasive impact" upon a person's life chances.[18] In particular, they shape the terms on which people can access the benefits of social cooperation, the development of their personal goals and aspirations, and the occasions on which they encounter the coercive power of the state. For many AI systems, this threshold is now also being met. In the words of legal scholar Rashida Richardson, AI systems are now being used to determine

> who will have their food subsidies terminated, how much healthcare benefits a person is entitled to, and who is likely to be a victim of crime …. [They] have concrete consequences for individuals and communities, such as increased law enforcement harassment, deportation, denial of housing or employment opportunities, and death.[19]

The stakes are therefore sufficiently high for principles of justice to be invoked.

If the preceding argument is correct, then it has a number of implications for the character of AI systems that are deployed in these spaces. These include:
   *Publicity*. The theory of justice developed by Rawls aims to identify principles for the governance of major institutions that can be justified to people despite variation in their beliefs about what a good or perfect society would look like. Situated in the "original position," people are asked to choose principles of justice for society from behind a "veil of ignorance," which prevents them from knowing the position in society they will occupy. Given that people are not able to tailor principles in a way that is prejudicial to their own interests, the principles selected are held to be fair and thus ones that people can willingly endorse. Moreover, given that society at times relies upon coercive sanctions to enforce norms via legal instruments, Rawls holds that "the grounds of its institutions should stand up to public scrutiny."[20] This "publicity condition" ensures that "citizens are in a po-sition to know and to accept the pervasive influences of the basic structure that shape their conception of themselves, their character and their ends."[21]

The publicity condition has important ramifications for the uses of AI that we have discussed. In particular, the requirement appears to sit in tension with elements of what Pasquale terms the "black box society," including the use of opaque hiring and credit allocation algorithms that shape citizen's life prospects.[22] Conversely, it helps to explain why calls for certain kinds of explanation are justified in the context of these AI systems: they are part of a more general entitlement citizens hold in relation to the institutions that shape their lives.[23] Moreover, as we have seen, mere knowledge of the principles that govern the behavior of public institutions is not sufficient to render them legitimate. People must also be in a position to accept the principles despite variation in personal moral beliefs. In the context of AI, this means that the integration and deployment of the technology must be justifiable in terms of an ideal of public reason.[24] It should, in the words of philosopher Jonathan Quong, be something that is acceptable "to each of us by reference to some common point of view, despite our deep differences and disagreements."[25]

One major consequence of this requirement is that an appeal to purely private goals, whether those of an individual or organization, will not be sufficient to justify the adoption or deployment of AI systems in certain public contexts. Instead, a public rationale must be provided. Second, the publicity condition points toward the existence of a derivative duty on the part of those who develop and deploy AI systems – to test them prior to deployment and to offer nontechnical explanations of their performance – so that the models are amenable to this kind of informed public debate, discussion, and evaluation.

*Basic liberties*. The first principle of justice endorsed by Rawls requires that "each person has the same indefeasible claim to a fully adequate system of basic liberties, which scheme is compatible with the same scheme of liberties for all."[26] These basic liberties work to "protect fundamental interests that have special significance" and include, at a minimum, "freedom of thought and liberty of conscience; the political liberties and freedom of association, as well as the freedoms specified by the liberty and integrity of the person; and finally, the rights and liberties covered by the rule of law."[27] The basic liberties are relevant for the design and deployment of AI systems in at least two respects.

The first concerns the protection they accord citizens. A major aim of this principle is to ground "a secure common status of equal citizenship" for society's members.[28] This aspiration dovetails effectively with the notion that institutions must be "effectively and impartially administered," given that deviation from this ideal contravenes the rights and liberties covered by the rule of law.[29] Understood in this way, the enjoyment of equal basic liberties stands in opposition to certain forms of algorithmic discrimination. As philosopher Tommie Shelby notes, the principle prohibits cases in which the rules of a public institution are applied un-evenly, including situations "where the administration or enforcement of its rules and procedures is frequently distorted by the racial prejudice and bias of its officials."[30] While the primary concern at the time of writing was with the bias of human officials, there is no reason to think that bias is less problematic when it is inherited by automated decision systems that perform a similar function. Indeed, given the potential for these systems to perform better than human decision-makers, one might think that the errors they make are more egregious.

Second, the list of basic liberties provided by Rawls is dynamic and varies according to the sociotechnical character of the society to which they apply. The initial list is based upon conditions that are held to be necessary for the development of moral autonomy and personhood irrespective of time or place (such as freedom of conscience). However, Rawls also notes that it is wise to take a "historical approach," which involves identifying additional rights that have demonstrable practical value for different societies at a specific moment in time. As a consequence, Rawls writes that "it is perhaps impossible to give a complete specification of these liberties independent from the particular circumstances – social, economic and *technological* – of a given society."[31] On each occasion, the key question is: what liberties are necessary to protect individuals in the development and pursuit of the conception of the good life, given the specific sociotechnical character of the society in which they live?

The potential for intrusion created by modern AI systems, both in terms of the data they are trained on and their ability to influence or foreshadow subsequent behavior, has given range to a host of new concerns.[32] To guard against these risks, it is quite possible that a right to privacy should now be added to the list of basic liberties. Although the grounds of a potential right to privacy are philosophically contested, legal scholar Andrei Marmor argues that they are closely connected to our well-being and are "violated when somebody manipulates, without adequate justification, the relevant environment in ways that significantly diminish your ability to control what aspects of yourself you reveal to others."[33] Given Rawls's concern with the ability of citizens to pursue a conception of the good life that is free from unwarranted interference, the basic liberties may now include protection against invasive forms of surveillance or behavioral manipulation.

*Fair equality of opportunity*. Rawls's second principle of justice holds that:

> Social and economic inequalities are to satisfy two conditions: first they are to be attached to offices and positions open to all under conditions of fair equality of opportunity; and second, they are to be to the greatest benefit of the least-advantaged member of society.

This principle also has far-reaching implications for AI. Starting with the first condition, it holds that fair equality of opportunity – not just formal equality of opportunity – must be achieved when determining how opportunities are allocat-ed between citizens. Thus, the requirements of justice are not met simply via the adoption of processes that do not discriminate against people on the basis of certain protected characteristics at the point at which a decision is made. Instead, a just society will aim to eliminate the impact of a wide range of unchosen features on their life prospects. The most natural reading of this requirement includes features such as a person's race, sex, class, and other contingencies of birth. Once the relevant adjustments have been made, we should arrive at a situation in which people of similar ability have roughly equal prospects of success.

In the context of debates around AI fairness, the implications of this principle are potentially significant. They mean moving away from a purely formal conception of fairness as equal treatment or "de-biasing" and thinking about how these tools can actively mitigate the effect of bias that exists at a societal level through various corrective measures.[34] As information scientists Solon Barocas and Andrew Selbst have noted, this debate mirrors a long-running discussion in jurisprudence about the appropriate goal of antidiscrimination legislation.[35] Whereas the anticlassification approach is concerned with equal treatment in a formal sense that involves eliminating unfairness that "individuals in certain protected classes experience due to decision makers' choices," antisubordination reaches beyond that and is more closely aligned with Rawls's fair equality of opportunity principle.[36] It holds that the goal of antidiscrimination law is "to eliminate status-based inequality due to membership in those classes, not as a matter of procedure, but of substance."[37] If this is the appropriate normative standard for AI systems performing key social functions, then we need further research and public discussion about what substantively fair outcomes look like in practice, and about how AI systems can support this societal objective.

*The difference principle*. The second condition, commonly known as the *difference principle*, also has implications for the design and deployment of AI. This principle holds that for institutional practices to be just, all inequalities in the distribution of "social primary goods" (which include income, wealth, and the "social bases of self-respect") must work to the greatest advantage of the least advantaged member of society. It follows that when AI is integrated into a key social practice, in a way that affects the overall distribution of benefits and burdens, it is pertinent to ask whether it does the most it possibly can do to improve the position of the least advantaged member of that system. This is a challenging question, and one that points toward a potentially exacting standard for AI deployment. Cumulatively, it redraws the scope of current debates about how to evaluate the impact of AI systems, making the impact of these systems on the distribution of wealth, resources, and social standing an important desideratum, while also proposing a standard for evaluation that is strongly egalitarian.

In terms of the practical implications of the difference principle, it seems clear that any technology that worsens the position of the most disadvantaged mem-ber of society in absolute terms, once it has been incorporated into relevant social practice, will fail to meet a key requirement of justice irrespective of other benefits it may bring (such as scalability or efficiency). Yet fully realized, the difference principle proposes a higher standard than simply improving the status quo: it suggests that the AI systems must make the worst-off as well-off as they can, relative to alternative system designs, or otherwise risk being part of a practice that is not fully legitimate. This standard is most clearly applicable to AI systems that have been integrated into core economic functions. However, it potentially has much wider applicability, extending to the full range of sociotechnical systems that shape a person's access to resources or impact upon their social standing and sense of self-worth.

Moreover, this demand is not met simply by the present combination of private innovation in the space of AI and *post hoc* economic redistribution. For while the redistribution of wealth is an important component of justice on any account, we also need to consider how sociotechnical systems influence the production of inequality *ex ante*. This is because there are likely to be opportunities to intervene at this point that do not arise later on. Indeed, given the emphasis Rawls places on self-esteem, in particular, there are opportunities to create fairer AI systems (that minimize inequalities in the first place) that cannot be addressed simply by making those who are badly off the *post facto* recipient of wealth transfers. Ultimately, these opportunities are what is missed when technology is bracketed out from liberal political theory: we may fail to consider an important site of distributive justice and hence mistakenly believe that society is substantively just when this is not the case and when impermissible forms of technologically induced inequality are hiding in plain sight.

These arguments are presented in the spirit of constructive co-investigation. My main purpose has been to illustrate the kind of rich moral insight that results from extending the domain of distributive justice to include AI systems. Clearly, more work needs to be done to substantiate these claims and translate them into guidelines for technologists and public officials. Indeed, as this preliminary account makes clear, it is possible that tensions will emerge, for example, between the notion derived from the liberty principle that individuals must be treated in a consistent manner and the notion, anchored in the fair equality of opportunity principle, that groups must experience similar outcomes.[38] Nonetheless, core elements of this approach seem destined to remain in place. If AI is, as I have argued, now part of the major sociotechnical practices that make up the basic structure of society, then its design and deployment should feed into practices that are amenable to public justification, support citizen's rights, and embody substantive properties connected with an egalitarian conception of justice. In these contexts, the appropriate goal of AI alignment is not an open ques-tion. Rather, the development and deployment of AI systems represent a new site for the operation of principles of distributive justice.[39]

I have argued that when AI is integrated into the functioning of major institutions and social practices, the norms that apply to the basic structure of society also apply to these systems. To ground this claim, I pointed to the role that AI now plays in augmenting or undermining background justice, and to a range of profound effects that AI has on the lives of citizens, particularly in the context of our major political and economic institutions. However, the preceding argument leaves open the question of alignment for AI systems deployed outside of key socioeconomic practices. In these environments, is it perhaps the prerogative of engineers or organizations to align AI systems with their own preferred values?

To answer this question, we need to understand how the two grounding conditions map onto other kinds of AI systems. Taking the profound effects condition first, it seems likely that many AI deployments meet this threshold. For example, AI-powered search and curation systems are deeply integrated into prevailing social epistemological practices, functioning as custodians for the legibility of the world around us, and influencing what we take to be true on an individual and collective level. Moreover, recommendation systems have the potential to influence the development of our moral character in certain ways, shaping self-perception, preferences, and desires, even as they learn to "give us what we want." Yet when it comes to background justice the case for an expansive reading is less clear. As we have seen, background justice is concerned with society's ability to reproduce itself over time in such a way that the conditions for meaningful consent are preserved. From this vantage point, certain forms of interpersonal exploitation and domination are clearly objectionable. The salient question for AI systems is whether there are other roles they play, beyond those considered, that also mandate corrective measures of this kind.

Given uncertainty on this point, efforts to extend principles of distributive justice to a wider set of AI systems are somewhat inconclusive. Yet even on a restrictive reading of the scope of these principles, two further points remain to be made. First, principles of distributive justice have implications for AI systems that are not part of the basic structure. On this point, Rawls notes that we should not regard the "political and the nonpolitical domains as two separate, disconnected spaces . . . each governed solely by its own distinct principles."[40] Instead, principles of justice place "essential restrictions" on all other activities. By way of illustration, Rawls does not consider the media to be part of the basic structure of society. However, requirements of justice nonetheless entail that this sphere of activity must be structured in a way that ensures the fair value of the political liberties. In the context of AI, it means that, at a minimum, public deployments of this technology must be compatible with principles of justice. Moreover, on an individual level, liberal political theory holds that we are all under a "duty of justice" to support the operation of institutions that enable cooperation on terms that are fair. When applied to groups concerned with the creation of new technologies, duties of justice plausibly become "duties of deployment" to support, and not subvert, the functioning of just institutions.

Second, the demand for public justification in the context of AI deployment may well extend beyond the basic structure. As social scientist Langdon Winner argues, when the impact of a technology is sufficiently great, this fact is, by itself, sufficient to generate a free-standing requirement that citizens be consulted and given an opportunity to influence decisions.[41] Absent such a right, citizens would cede too much control over the future to private actors, something that sits in tension with the idea that they are free and equal. Against this claim, it might be objected that it extends the domain of political justification too far, in a way that risks crowding out room for private experimentation, exploration, and the development of projects by citizens and organizations. However, the objection rests upon the mistaken view that autonomy is promoted by restricting the scope of justificatory practices to as narrow a subject matter as possible. In reality, this is not the case: what matters for individual liberty is that practices that have the potential to interfere with this freedom are appropriately regulated so that infractions do not come about. Understood in this way, the demand for public justification stands in opposition not to personal freedom but to forms of unjust technological imposition.[42]

The demand for justice in the context of AI is well-founded. Considered through the lens of distributive justice, key principles that govern the fair organization of our social, political, and economic institutions also apply to AI systems that are embedded in these practices. One major consequence of this is that liberal and egalitarian norms of justice apply to AI tools and services across a range of contexts. When they are integrated into society's basic structure, these technologies should, I have argued, support citizens' basic liberties, promote fair equality of opportunity, and provide the greatest benefit to those who are worst-off. Moreover, deployments of AI outside of the basic structure must still be compatible with the institutions and values that justice requires. There will always be valid reasons, therefore, to consider the relationship of technology to justice when it comes to the deployment of AI systems.

AUTHOR'S NOTE

I would like to thank Laura Weidinger, William Isaac, Julia Haas, Conor Griffin, Sean Legassick, Christopher Summerfield, Allan Dafoe, Shakir Mohamed, Brittany Smith, Courtney Biles, Aliya Ahmad, Geoff Keeling, Thomas K Gilbert, Abeba Birhane, Jeff Howard, Juri Viehoff, Johannes Himmelreich, James Manyika, and the editorial team at *Dædalus* for their support with this work.

ABOUT THE AUTHOR

**Iason Gabriel** is a Staff Research Scientist at DeepMind. He has published in such journals as *Minds and Machines*, *The Philosophical Quarterly*, and *The Journal of Applied Philosophy*.

ENDNOTES